\documentclass[journal]{IEEEtran}
\usepackage{amssymb}
\usepackage{amsmath}
\usepackage{amsfonts}
\usepackage{algorithm}
\usepackage{algorithmic}
\usepackage{graphicx}

\newtheorem{theorem}{Theorem}

\newtheorem{definition}[theorem]{Definition}
\newtheorem{example}[theorem]{Example}

\newtheorem{problem}[theorem]{Problem}
\newtheorem{proposition}[theorem]{Proposition}
\newtheorem{remark}[theorem]{Remark}

\numberwithin{equation}{section}
\numberwithin{equation}{subsection}
\numberwithin{theorem}{subsection}

\input{tcilatex}
\begin{document}

\title{The Incidence and Cross Methods for Efficient Radar Detection\medskip 
}
\author{{\Large Alexander Fish and Shamgar Gurevich}\thanks{%
This material is based upon work supported by the Defense Advanced Research
Projects Agency (DARPA) award number N66001-13-1-4052. Any opinions,
findings, and conclusions or recommendations expressed in this publication
are those of the authors and do not necessarily reflect the views of DARPA.
This work was also supported in part by NSF Grant DMS-1101660 - "The
Heisenberg--Weil Symmetries, their Geometrization and
Applications".\smallskip {}}\thanks{%
A. Fish is with the School of Mathematics and Statistics, University of
Sydney, Sydney, NSW 2006, Australia. Email:
alexander.fish@sydney.edu.au.\smallskip\ } \thanks{%
S. Gurevich is with the Department of Mathematics, University of Wisconsin,
Madison, WI 53706, USA. Email: shamgar@math.wisc.edu. \medskip } \thanks{%
Accepted for publication in Proceedings of Allerton Conference on
Communication, Control, and Computing, October, 2013. }}
\maketitle

\begin{abstract}
The designation of the radar system is to detect the position and velocity
of targets around us. The radar transmits a waveform, which is reflected
back from the targets, and echo waveform is received. In a commonly used
model, the echo is a sum of a superposition of several delay-Doppler shifts
of the transmitted waveform, and a noise component. The delay and Doppler
parameters encode, respectively, the distances, and relative velocities,
between the targets and the radar. Using standard digital-to-analog and
sampling techniques, the estimation task of the delay-Doppler parameters,
which involves waveforms, is reduced to a problem for complex sequences of
finite length $N.$ In these notes we introduce the \textit{Incidence }and 
\textit{Cross }methods for radar detection. One of their advantages, is
robustness to inhomogeneous radar scene, i.e., for sensing small targets in
the vicinity of large objects. The arithmetic complexity of the incidence
and cross methods is $O(N\log N+r^{3})$ and $O(N\log N+r^{2}),$ for $r$
targets, respectively. In the case of noisy environment, these are the
fastest radar detection techniques. Both methods employ chirp sequences,
which are commonly used by radar systems, and hence are attractive for real
world applications.
\end{abstract}


\markboth{{\it The Incidence and Cross Methods for Efficient Radar Detection
--- By Alexander Fish and Shamgar Gurevich}}{}

\section{\textbf{Introduction\label{In}}}

\begin{figure}[ht]
\includegraphics[clip,height=5cm]{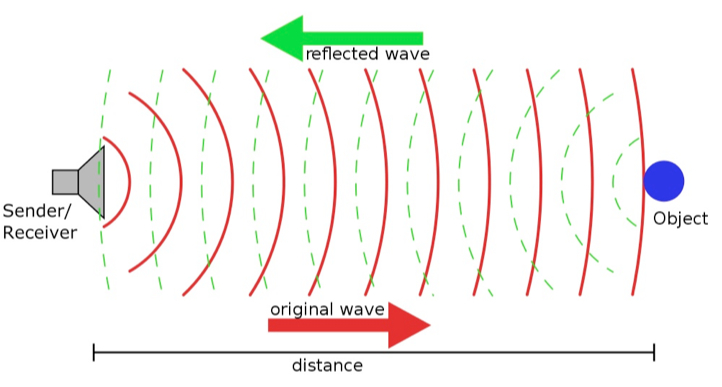}\\
\caption{Radar transmits waveform (red)
and received echo (green).}
\label{radar}
\end{figure}

The radar system provides detection of the position and velocity of moving
targets. The radar transmits---see Figure \ref{radar} for illustration---an
analog waveform $S_{A}(t),$ $t\in 
\mathbb{R}
,$ of bandwidth $W$. While the actual waveform is modulated onto a carrier
frequency $f_{c}\gg W,$ we consider a widely used complex baseband model for
the multi-target channel (see Section I.A. in \cite{FGHSS} and references
therein). In addition, we make the sparsity assumption on the finiteness of
the number of targets. The waveform $S_{A}$ hits the targets, and the analog
waveform received as echo is given by\footnote{%
In this paper $i$ $=\sqrt{-1}.$}%
\begin{equation}
R_{A}(t)=\sum_{k=1}^{r}\beta _{k}\cdot \exp (2\pi if_{k}t)\cdot
S_{A}(t-t_{k})+\mathcal{W(}t),  \label{RA}
\end{equation}%
where $r$---called the \textit{sparsity }of the channel---denotes the number
of targets, $\beta _{k}\in 
\mathbb{C}
$ is the \textit{attenuation coefficient}, $f_{k}\in 
\mathbb{R}
$ is the \textit{Doppler shift, }$t_{k}\in 
\mathbb{R}
_{+}$ is the \textit{delay} associated with the $k$-th target, and $\mathcal{%
W}$ denotes a random white noise. We assume the normalization $%
\sum_{k=1}^{r}\left\vert \beta _{k}\right\vert ^{2}\leq 1$. The Doppler
shift depends on the relative velocity, and the delay encodes the distance,
between the radar and the target. We will call 
\begin{equation}
(t_{k},f_{k}),\text{ }k=1,...,r,  \label{CP}
\end{equation}%
\textit{channel parameters, }and the main objective of radar detection is:

\begin{problem}[\textbf{Analog Radar Detection}]
Estimate the parameters $(\ref{CP}\bigskip ).$
\end{problem}

\subsection{\textbf{Digital Radar Detection\label{DRD}}}

Using standard digital-to-analog and sampling techniques (see Section I.A.
in \cite{FGHSS} and references therein), the estimation task, which involves
waveforms, is reduced to the following problem for complex sequences of
finite length $N.$ We consider the set of integers $%
\mathbb{Z}
_{N}=\{0,1,...,N-1\}$ with addition and multiplication modulo $N.$ For the
rest of the paper we assume that $N$ is an odd prime. We will denote by 
\begin{equation*}
\mathcal{H=\{}S:%
\mathbb{Z}
_{N}\rightarrow 
\mathbb{C}
\},
\end{equation*}%
the vector space of complex valued functions on $%
\mathbb{Z}
_{N}$, and we refer to it as the \textit{Hilbert space of sequences. }We
define the\textit{\ channel operator} $H$ acting on $S\in \mathcal{H}$ by%
\footnote{%
We define $e(t)=\exp (2\pi it/N),$ $t\in 
\mathbb{R}
.$} 
\begin{equation}
H(S)[n]=\sum_{k=1}^{r}\alpha _{k}\cdot e(\omega _{k}n)\cdot S[n-\tau _{k}],%
\text{ }n\in 
\mathbb{Z}
_{N},  \label{H}
\end{equation}%
with $\alpha _{k}\in 
\mathbb{C}
$\textit{,} $\sum_{k=1}^{r}\left\vert \alpha _{k}\right\vert ^{2}\leq 1,$
and $\tau _{k},\omega _{k}\in 
\mathbb{Z}
_{N}.$ In particular, for every transmitted sequence $S\in \mathcal{H}$ we
have the associated received sequence%
\begin{equation}
R[n]=H(S)[n]+\mathcal{W}[n],\text{ }n\in 
\mathbb{Z}
_{N},  \label{R}
\end{equation}%
where $\mathcal{W}\in \mathcal{H}$ denotes a random white noise. For the
rest of these notes we assume that all the coordinates of the sequence $%
\mathcal{W}$ are independent, identically distributed random variables of
expectation zero. In analogy with the physical channel model described by
Equation (\ref{RA}), we will call $\alpha _{k},\tau _{k},\omega _{k},$ $%
k=1,...,r,$ \textit{attenuation coefficients}, \textit{delays}, and \textit{%
Doppler shifts}, respectively.

\begin{problem}[\textbf{Digital Radar Detection}]
\label{DRDP}\bigskip Design $S\in \mathcal{H}$ (or small family of
sequences), and effective method to extract the channel parameters
\end{problem}

\begin{equation}
(\tau _{k},\omega _{k}),\ k=1,...,r,  \label{DCP}
\end{equation}%
using $S$ and $R$ satisfying (\ref{R}).\medskip

\begin{remark}
\label{RDRD}The relation between the physical (\ref{CP}) and the discrete (%
\ref{DCP}) channel parameters is as follows (see Section I.A. in \cite{FGHSS}
and references therein): If a standard method suggested by sampling theorem
is used for the discretization, and $S_{A}$ has bandwidth $W$, then $\tau
_{k}=t_{k}W$ $\func{mod}$ $N$, and $\omega _{k}=Nf_{k}/W$ $\func{mod}$ $N$,
provided that $t_{k}\in \frac{1}{W}\mathbb{Z},$ and $f_{k}\in \frac{W}{N}%
\mathbb{Z},$ $k=1,...,r.$ In particular, we note that the integer $N$
determines the frequency resolution of the radar detection, i.e., the
resolution is of order $W/N.$
\end{remark}

\subsection{\textbf{Ambiguity Function and Pseudo-Random Method}}

A classical method to compute the channel parameters (\ref{DCP}) is the 
\textit{pseudo-random method }\cite{GG, GHS, HCM, TV, V, WG}. It uses two
ingredients - the ambiguity function, and a pseudo-random sequence.\medskip

\subsubsection{\textbf{Ambiguity Function}}

In order to reduce the noise component in (\ref{R}), it is common to use the
ambiguity function that we are going to describe now. \ The space $\mathcal{H%
}$ is equipped with the standard inner product%
\begin{equation*}
\left\langle f,g\right\rangle =\sum_{n=0}^{N-1}f[n]\overline{g}[n],\text{ \ }%
f,g\in \mathcal{H}\text{,}
\end{equation*}%
where $\overline{g}$ denotes the complex conjugate of $g.$ In addition, we
consider the \textit{Heisenberg operators }$\pi (\tau ,\omega ),$ $\tau
,\omega \in 
\mathbb{Z}
_{N},$ which act on $f\in \mathcal{H}$ by%
\begin{equation}
\left[ \pi (\tau ,\omega )f\right] [n]=e(-2^{-1}\tau \omega )\cdot e(\omega
n)\cdot f[n-\tau ],  \label{HO}
\end{equation}%
where $2^{-1}$ denotes $(N+1)/2,$ the inverse of $2$ $\func{mod}$ $N.$
Finally, the \textit{ambiguity function }of two sequences $f,g\in \mathcal{H}
$ is defined\footnote{%
For our purposes it will be convenient to use this definition of the
ambiguity function. The standard definition appearing in the literature is $%
A(f,g)[\tau ,\omega ]=\left\langle e(\omega n)f(n-\tau ),g[n]\right\rangle .$%
} as the $N\times N$ matrix 
\begin{equation}
\mathcal{A(}f,g)[\tau ,\omega ]=\left\langle \pi (\tau ,\omega
)f,g\right\rangle ,\text{ \ }\tau ,\omega \in 
\mathbb{Z}
_{N}.  \label{AF}
\end{equation}

\begin{remark}[\textbf{Fast Computation of Ambiguity Function}]
\label{FC}The restriction of the ambiguity function to a line in the
delay-Doppler plane, can be computed in $O(N\log N)$ arithmetic operations
using fast Fourier transform \cite{R}. For more details, including explicit
formulas, see Section V of \cite{FGHSS}. Overall, we can compute the entire
ambiguity function in $O(N^{2}\log N)$ operations.\medskip
\end{remark}

For $R$ and $S$ satisfying (\ref{R}), the law of the iterated logarithm
implies that, with probability going to one, as $N$ goes to infinity, we
have 
\begin{equation}
\mathcal{A}(S,R)[\tau ,\omega ]=\mathcal{A}(S,H(S))[\tau ,\omega
]+\varepsilon _{N},  \label{noise}
\end{equation}%
where $\left\vert \varepsilon _{N}\right\vert \leq \sqrt{2\log \log N}/\sqrt{%
N\cdot SNR},$ with $SNR$ denotes the \textit{signal-to-noise ratio}\footnote{%
We define $SNR=\left\langle S,S\right\rangle /\left\langle \mathcal{W},%
\mathcal{W}\right\rangle $.}.

\subsubsection{\textbf{Pseudo-Random Sequences}}

\begin{figure}[ht]
\includegraphics[clip,height=5cm]{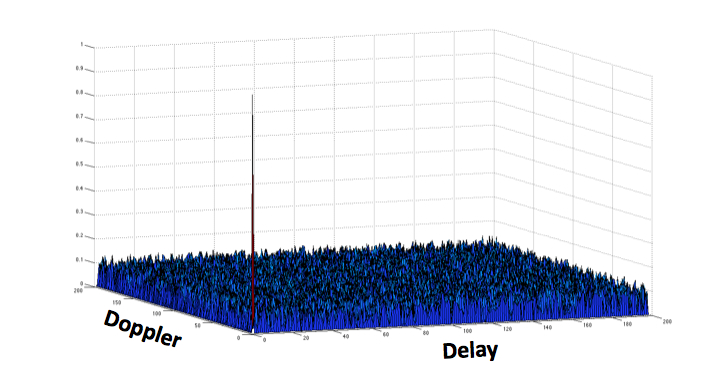}\\
\caption{Profile of $\mathcal{A}(\protect%
\varphi ,\protect\varphi )$ for $\protect\varphi $ PR.}
\label{PR}
\end{figure}

\bigskip We will say that a norm-one sequence $\varphi \in \mathcal{H}$ is $%
B $-\textit{pseudo-random, }$B\in 
\mathbb{R}
$\textit{---}see Figure \ref{PR} for illustration---if for every $\left(
\tau ,\omega \right) \neq (0,0)$ we have \textit{\ }%
\begin{equation}
\left\vert \mathcal{A}(\varphi ,\varphi )[\tau ,\omega ]\right\vert \leq B/%
\sqrt{N},  \label{pr}
\end{equation}%
There are several constructions of families of pseudo-random (PR) sequences
in the literature (see \cite{GG, GHS, WG} and references therein).

\subsubsection{\textbf{Pseudo-Random Method}}

\begin{figure}[ht]
\includegraphics[clip,height=5cm]{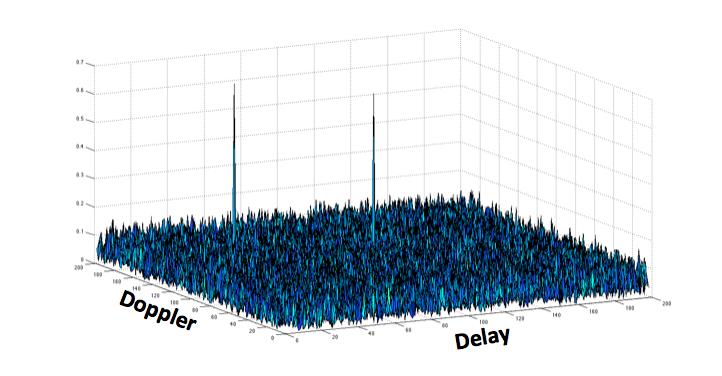}\\
\caption{Profile of $\mathcal{A}(\protect\varphi ,H(\protect\varphi ))$, for $%
\protect\varphi $ PR, $N=199,$ and channel parameters $(50,150)$, $%
(100,100), $ $(150,50),$ with attenuation coefficients $0.7,$ $0.7,$ $0.1,$
respectively. Note the insensibility of the channel parameter $(150,50)$
associated with the small attenuation coefficient.  }
\label{PRM}
\end{figure}

Consider a pseudo-random sequence $\varphi $, and assume for simplicity that 
$B=1$ in (\ref{pr}). Then---see Figure \ref{PRM} for illustration---we have 
\begin{eqnarray}
&&\mathcal{A}(\varphi ,H(\varphi ))[\tau ,\omega ]  \label{prm} \\
&=&\left\{ 
\begin{array}{c}
\widetilde{\alpha }_{k}+\tsum\limits_{j\neq k}\widetilde{\alpha }_{j}/\sqrt{N%
},\text{ \ if }\left( \tau ,\omega \right) =\left( \tau _{k},\omega
_{k}\right) ,\text{ }1\leq k\leq r; \\ 
\tsum\limits_{j}\widehat{\alpha }_{j}/\sqrt{N},\text{ \ \ \ \ \ otherwise, \
\ \ \ \ \ \ \ \ \ \ \ \ \ \ \ \ \ \ \ \ \ \ \ \ \ }%
\end{array}%
\right.  \notag
\end{eqnarray}%
where $\widetilde{\alpha }_{j},$ $\widehat{\alpha }_{j},$ $1\leq j\leq r,$
are certain multiples of the $\alpha _{j}$'s by complex numbers of absolute
value less or equal to one. In particular, we can compute the delay-Doppler
parameter $\left( \tau _{k},\omega _{k}\right) $ if the associated
attenuation coefficient $\alpha _{k}$ is sufficiently large with respect to
the others. It appears---see Figure \ref{PRM} for illustration---as a peak
of $\mathcal{A}(\varphi ,H(\varphi )).$

\medskip


\begin{remark}
\label{RPR}We have

\begin{enumerate}
\item \textbf{Arithmetic Complexity. }The arithmetic complexity of the
pseudo-random method is $O(N^{2}\log N),$ using Remark \ref{FC}. \medskip

\item \textbf{Large Deviation of Attenuation Coefficients. }From Identity (%
\ref{prm}), we deduce that the pseudo-random method will fail to detect
delay-Doppler parameter $\left( \tau _{k},\omega _{k}\right) $ associated
with attenuation coefficient $\alpha _{k}$ which is small in magnitude
compare to $\tsum \widetilde{\alpha }_{j}/\sqrt{N}.$ We illustrate this in
Figure \ref{PRM}, where the channel parameter $(150,50)$ can not be detected
because it is associated with the small attenuation coefficient equal to $%
0.1.\medskip $

\item \textbf{Noise. }From (\ref{noise}) we conclude that, a target is
detectable by the pseudo-random method only if the associated attenuation
coefficient is of magnitude larger than $\sqrt{2\log \log N}/\sqrt{N\cdot SNR%
}.$
\end{enumerate}
\end{remark}

\subsection{\textbf{Arithmetic Complexity Problem}}

For applications to sensing, that require sufficiently high frequency
resolution, we will need to use sequences of large length $N$ (see Remark %
\ref{RDRD}). In this case, the arithmetic complexity $O(N^{2}\log N)$ of the
pseudo-random method might be too high. Note that to compute one entry of
the ambiguity function already takes $N$ operations.\medskip

\begin{problem}[\textbf{Arithmetic Complexity}]
Solve Problem \ref{DRDP}, with method for extracting the channel parameters (%
\ref{DCP}), which requires almost linear arithmetic complexity.
\end{problem}\medskip

\begin{figure}[ht]
\includegraphics[clip,height=5cm]{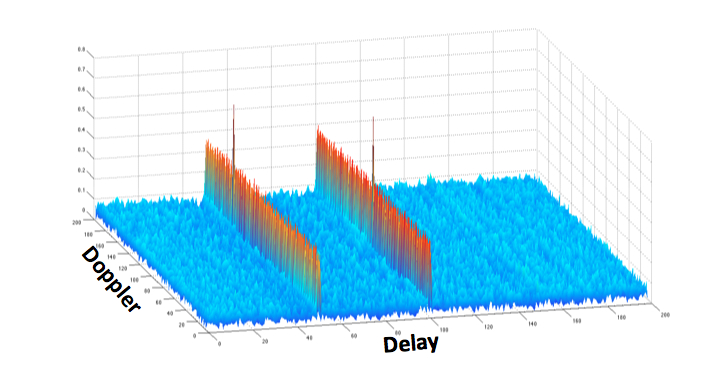}\\
\caption{Profile of $\mathcal{A(}%
f_{L},H(f_{L}))$ for flag $f_{L},$ $L=\{(0,\protect\omega )\},$ $N=199,$ and
channel parameters $(50,150)$, $(100,100),$ with attenuation
coefficients $0.7,$ $0.7,$ respectively.  }
\label{FlagM}
\end{figure}


In \cite{FGHSS} the flag
method was introduced in order to deal with the complexity problem. It
computes $r$ channel parameters in $O(rN\log N)$ arithmetic operations. For
a given line $L$ in the plane $%
\mathbb{Z}
_{N}\times 
\mathbb{Z}
_{N},$ one construct a sequence $f_{L}$---called flag---with ambiguity
function $\mathcal{A(}f_{L},H(f_{L}))$ having special profile---see Figure %
\ref{FlagM} for illustration. It is essentially supported on shifted lines
parallel to $L$, that pass through the delay-Doppler shifts (\ref{DCP}), and
have peaks there. This suggests a simple algorithm to extract the channel
parameters. First compute $\mathcal{A(}f_{L},H(f_{L}))$ on a line $M$
transversal to $L,$ and find the shifted lines on which $\mathcal{A(}%
f_{L},H(f_{L}))$ is supported. Then compute $\mathcal{A(}f_{L},H(f_{L}))$ on
each of the shifted lines and find the peaks. The overall complexity of the
flag algorithm is therefore $O(rN\log N)$, using Remark \ref{FC}.\bigskip

In these notes we suggest radar detection methods, that, in the multi-target
regime, have much better arithmetic complexity.

\subsection{\textbf{Inhomogeneous Radar Scene Problem}}

We would like to estimate the channel parameters (\ref{DCP}) also in the
case of large deviation of attenuation coefficients (see Remark \ref{RPR}).
This task arises in \textit{inhomogeneous radar scene}, i.e., in attempt to
sense small targets in the vicinity of large objects---see Figure \ref%
{caradar} for illustration.

\begin{figure}[ht]
\includegraphics[clip,height=5cm]{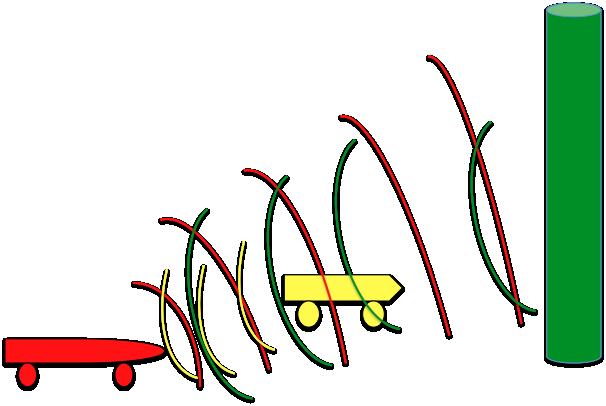}\\
\caption{Car
radar (red) transmits waveform, and receives echo from car (yellow) and
tower (green).  }
\label{caradar}
\end{figure}


\begin{problem}[\textbf{Inhomogeneous Radar Scene}]
\label{IS}Solve Problem \ref{DRDP} in the case of large deviation of
attenuation coefficients.
\end{problem}
\medskip

\begin{remark}
We note that both the pseudo-random and flag methods \cite{FGHSS} use
pseudo-random sequences. Hence, their applicability for inhomogeneous radar
scene is limited (see Figure \ref{PRM}). 
\end{remark}

\subsection{\textbf{Solutions to the Arithmetic Complexity and Inhomogeneous
Radar Scene Problems}}

In these notes we introduce the Incidence and Cross methods, which estimate
the channel parameters in complexity of $O(N\log N+r^{3}),$ and $O(N\log
N+r^{2})$, respectively. This is a striking improvement over the flag and
pseudo-random methods, in the realistic sparsity regime $r\ll N.$ In
addition, the incidence and cross methods suggest solutions to the
inhomogeneous radar scene problem. Both methods use the special sequences
called \textit{chirps.} This makes them attractive for real world
applications, since the commonly used Linear Frequency Modulated (LFM)\
radar employs chirp sequences.\medskip

\begin{remark}
A more comprehensive treatment of the incidence and cross methods, including
further development, statistical analysis, and proofs, will appear elsewhere.
\end{remark}

\section{\textbf{Chirp Sequences\label{CS} }}

In this section we introduce chirp sequences, and discuss their correlation
properties. In addition, we recall their eigenfunction property for a
certain commuting family of Heisenberg operators.

\subsection{\textbf{Definition of the Chirp Sequences}}

\begin{figure}[ht]
\includegraphics[clip,height=5cm]{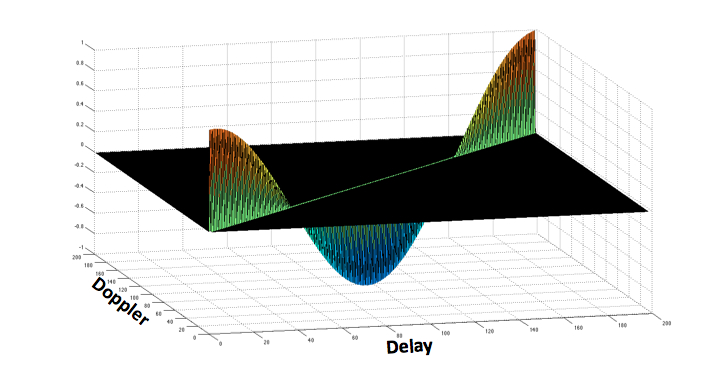}\\
\caption{Plot (real part) of $\mathcal{A}%
(C_{L_{1,1}},C_{L_{1,1}}),$ for chirp $C_{L_{1,1}}[n]=e[2^{-1}n^{2}-n]$,
associated with line $L_{1}=\{(\protect\tau ,\protect\tau )\}.$ }
\label{ChirpD}
\end{figure}

We have $N+1$ lines\footnote{%
In this paper by a \textit{line} $L\subset V$, we mean a line through $%
(0,0). $ In addition, by a \textit{shifted line, }we mean a subset of $V$ of
the form $L+v$, \ where $L$ is a line and $v\in V.$} in the \textit{discrete
delay-Doppler plane }$V=%
\mathbb{Z}
_{N}\times 
\mathbb{Z}
_{N}.$ For each $a\in 
\mathbb{Z}
_{N}$ we have the line $L_{a}=\left\{ (\tau ,a\tau );\tau \in 
\mathbb{Z}
_{N}\right\} $ of finite slope $a,$ and in addition we have the line of
infinite slope $L_{\infty }=\left\{ (0,\omega );\text{ }\omega \in 
\mathbb{Z}
_{N}\right\} .$ We define the orthonormal basis for $\mathcal{H}$ of \textit{%
chirp sequences} associated with $L_{a}$%
\begin{equation*}
\mathcal{B}_{L_{a}}=\left\{ C_{L_{a,b}}\text{; }b\text{ }\in 
\mathbb{Z}
_{N}\right\} ,
\end{equation*}%
where 
\begin{equation*}
C_{L_{a,b}}[n]=e(2^{-1}an^{2}-bn)/\sqrt{N},n\in 
\mathbb{Z}
_{N}.
\end{equation*}%
In addition, we have the orthonormal basis of chirp sequences associated
with $L_{\infty }$ 
\begin{equation*}
\mathcal{B}_{L_{\infty }}=\left\{ C_{L_{_{\infty },b}}\text{ ; }b\text{ }\in 
\mathbb{Z}
_{N}\right\} ,
\end{equation*}%
where 
\begin{equation*}
C_{L_{_{\infty },b}}=\delta _{b},
\end{equation*}%
denotes the Dirac delta sequence supported at $b.$ The chirp sequences
satisfy---see Figure \ref{ChirpD} for illustration---the following
properties:\medskip

\begin{theorem}[\textbf{Correlations}]
\label{CT}We have

\begin{enumerate}
\item \textit{Auto-correlation. }For every\textit{\ }$a,b\in 
\mathbb{Z}
_{N}$ 
\begin{equation*}
\mathcal{A}(C_{L_{a,b}},C_{L_{a,b}})[v]=\left\{ 
\begin{array}{c}
e(b\tau )\text{ \ if }v=(\tau ,a\tau ); \\ 
0\text{ \ \ \ \ \ if }v\notin L_{a}.\text{ \ }%
\end{array}%
\right.
\end{equation*}%
In addition, for every $b\in 
\mathbb{Z}
_{N}$ 
\begin{equation*}
\mathcal{A}(C_{L_{\infty ,b}},C_{L_{\infty ,b}})[v]=\left\{ 
\begin{array}{c}
e(b\omega )\text{ \ if }v=(0,\omega ); \\ 
\text{ \ }0\text{ \ \ \ \ if }v\notin L_{\infty }.\text{ \ }%
\end{array}%
\right. .
\end{equation*}

\item \textit{Cross-correlation. }For every two lines $L\neq M\subset V$,
and every $C_{L}\in \mathcal{B}_{L},C_{M}\in \mathcal{B}_{M},$%
\begin{equation*}
\left\vert \mathcal{A(}C_{L},C_{M})[v]\right\vert =1/\sqrt{N},
\end{equation*}%
for every $v\in V.\smallskip $
\end{enumerate}
\end{theorem}

\subsection{\textbf{Chirps as Eigenfunctions of Heisenberg Operators\label{E}%
}}

The Heisenberg operators (\ref{HO}) satisfy the commutation relations 
\begin{equation}
\pi (\tau ,\omega )\pi (\tau ^{\prime },\omega ^{\prime })=e(\omega \tau
^{\prime }-\tau \omega ^{\prime })\cdot \pi (\tau ^{\prime },\omega ^{\prime
})\pi (\tau ,\omega ),  \label{C}
\end{equation}%
for every $(\tau ,\omega ),(\tau ^{\prime },\omega ^{\prime })\in V.$ In
particular, for a given line $L\subset V,$ we have the family of commuting
operators $\pi (l),$ $l\in L.$ Hence they admit an orthonormal basis for $%
\mathcal{H}$ of common eigenfunctions. Important property of the chirp
sequences is that $\mathcal{B}_{L}$ is such a basis of eigenfunctions.
Indeed, it is easy to check that for every $a,b\in 
\mathbb{Z}
_{N}$%
\begin{equation}
\pi (\tau ,a\tau )C_{L_{a,b}}=e(b\tau )C_{L_{a,b}},\text{ }\tau \in 
\mathbb{Z}
_{N},  \label{E1}
\end{equation}%
and in addition for every $b\in 
\mathbb{Z}
_{N}$ 
\begin{equation}
\pi (0,\omega )C_{L_{\infty ,b}}=e(b\omega )C_{L_{\infty ,b}},\text{ \ }%
\omega \in 
\mathbb{Z}
_{N}.  \label{E2}
\end{equation}

\begin{remark}
\bigskip A function\footnote{%
We denote by $%
\mathbb{C}
^{\ast }$ the set of non-zero complex numbers.} $\psi :L\rightarrow 
\mathbb{C}
^{\ast }$, where $L\subset V$ is a line, is called \textit{character} if $%
\psi (l+l^{\prime })=\psi (l)\psi (l^{\prime })$, for every $l,l^{\prime
}\in L.$ Note that the functions $\psi _{a,b}:L_{a}\rightarrow 
\mathbb{C}
^{\ast },$ $\psi _{a,b}(\tau ,a\tau )=$ $e(b\tau ),$ and $\psi _{\infty
,b}:L_{\infty }\rightarrow 
\mathbb{C}
^{\ast },$ $\psi _{\infty ,b}(0,\omega )=$ $e(b\omega ),$ are characters, of 
$L_{a},$ and $L_{\infty },$ respectively. Sometimes, we will write equations
(\ref{E1}), (\ref{E2}) in one compact form 
\begin{equation*}
\pi (l)C_{L}=\psi _{L}(l)C_{L},\text{ }l\in L,\text{ }
\end{equation*}%
where $C_{L}\in \mathcal{B}_{L\text{ }}$is a chirp, and $\psi _{L}$ is a
character.
\end{remark}

\section{\textbf{Radar Detection using Chirps and the Matching Problem\label%
{RDMP}}}

One of the reasons to use chirps for radar detection, is their advantage in
the case of inhomogeneous radar scene. Let us elaborate on this. We define
the \textit{support }of the channel operator $H$ (\ref{H}) to be the set $%
supp(H)=\{\left( \tau _{k},\omega _{k}\right) ;$ $k=1,...,r\}.\medskip $

\begin{definition}[$L$\textbf{-genericity}]
Let $L\subset V$ be a line.

\begin{enumerate}
\item We say that a subset $X\subset V$ is \textit{generic }with respect to $%
L,$ if for every $u,v\in X$ we have $u-v\notin L.\medskip $

\item We say that the channel operator $H$ is $L$-\textit{generic, }if $%
supp(H)$ is generic with respect to $L$.\medskip
\end{enumerate}
\end{definition}

\begin{proposition}[\textbf{Genericity}]
\label{L-G}The probability $P,$ that a subset $\left\{
v_{1},...,v_{r}\right\} \subset V$ is generic with respect to a randomly
chosen line $L\subset V$, satisfies 
\begin{equation*}
P\geq 1-\frac{r^{2}-r}{2(N+1)}.
\end{equation*}
\end{proposition}

\begin{remark}
\label{rG}It follows from Proposition \ref{L-G}, that in the case $r\ll N^{%
\frac{1}{2}}$, the channel operator is generic, with high probability, with
respect to a randomly chosen line. \medskip\ 
\end{remark}

Assume that $H$ is $L$-generic. Then by Theorem \ref{CT}, for a chirp $%
C_{L}\in \mathcal{B}_{L}$ we have 
\begin{eqnarray}
&&\left\vert \mathcal{A}(C_{L},H(C_{L}))[v]\right\vert  \label{AL} \\
&=&\left\{ 
\begin{array}{c}
\left\vert \alpha _{k}\right\vert \text{ \ if }v\in L+(\tau _{k},\omega
_{k}),\text{ }k=1,...,r; \\ 
\text{ \ \ }0\text{ \ \ \ \ \ \ \ \ \ \ \ \ \ otherwise, \ \ \ \ \ \ \ \ \ \
\ \ \ \ \ \ \ }%
\end{array}%
\right.  \notag
\end{eqnarray}%
where $L+(\tau _{k},\omega _{k})$ denotes the shifted line $\left\{ l+(\tau
_{k},\omega _{k});\ l\in L\right\} .$

\begin{remark}
\bigskip The meaning of Identity (\ref{AL}) is that using a chirp we can
sense targets associated with small attenuation coefficients.\medskip\ 
\end{remark}

\begin{figure}[ht]
\includegraphics[clip,height=5cm]{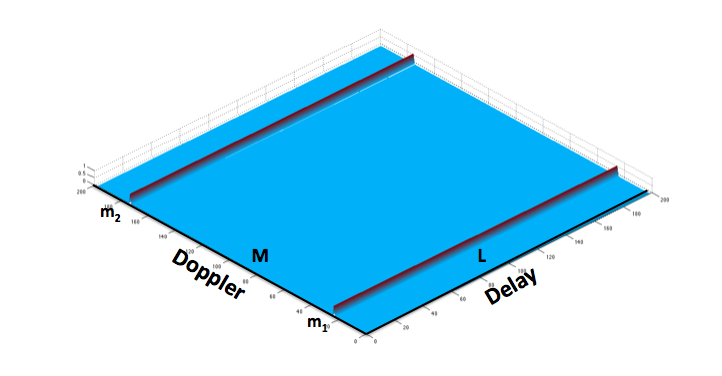}\\
\caption{$\left\vert \mathcal{A(}%
C_{L},H(C_{L}\mathcal{)})\right\vert $\textbf{\ }with\textbf{\ }$L=\{(%
\protect\tau ,0)\},$ $r=2,$ and\textbf{\ }$supp(H)=\{l_{1}+m_{2},$ $%
l_{2}+m_{1}\}.$ }
\label{ChirpL}
\end{figure}

Suppose we have an additional line $M\neq L,$ such that $H$ is $M$-generic.
In particular---see Figure \ref{ChirpL} for illustration---computing $%
\mathcal{A}(C_{L},H(C_{L}))$ on $M,$ we obtain $r$ peaks at points 
\begin{equation}
m_{1},...,m_{r}\in M.  \label{ms}
\end{equation}%

\begin{figure}[ht]
\includegraphics[clip,height=5cm]{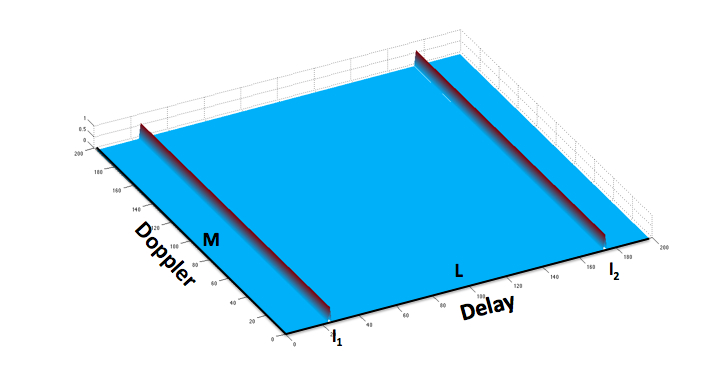}\\
\caption{$\left\vert \mathcal{A(}C_{M},H(C_{M}\mathcal{)})\right\vert $
with $M=\{(0,\protect\omega )\},$ $r=2,$ and $supp(H)=\{l_{1}+m_{2},$ $%
l_{2}+m_{1}\}.$  }
\label{ChirpM}
\end{figure}


In the same way---see Figure \ref{ChirpM} for illustration---computing $%
\mathcal{A}(C_{M},H(C_{M}))$ on $L,$ we obtain $r$ peaks at points 
\begin{equation}
l_{1},...,l_{r}\in L.  \label{ls}
\end{equation}%
Note that every channel parameter (\ref{DCP}) is represented uniquely by a
suitable $l_{i}+m_{j}$, for some $i,j$. \medskip\ 

\begin{problem}[\textbf{Matching}]
\label{M}Find the $r$ points from $l_{i}+m_{j}$, $1\leq i,j\leq r,$ which
belong to $supp(H).\medskip $
\end{problem}

In these notes we will propose resolutions to the matching problem, that
will be efficient in terms of arithmetic complexity, and work well also if
we add to $H$ a reasonable noise sequence $\mathcal{W}$ as in (\ref{R}).

\section{\textbf{The Incidence Method\label{TIM}}}

The \textit{incidence method} is the first resolution for the matching
problem that we discuss. The main idea already appears in various places in
the literature (see Figure 29 on page 307 of \cite{L}). Our contribution is
a low arithmetic complexity implementation of this method, and certain
mathematical analysis of its performance.

\subsection{\textbf{The Incidence Method\label{IM}}}

\begin{figure}[ht]
\includegraphics[clip,height=5cm]{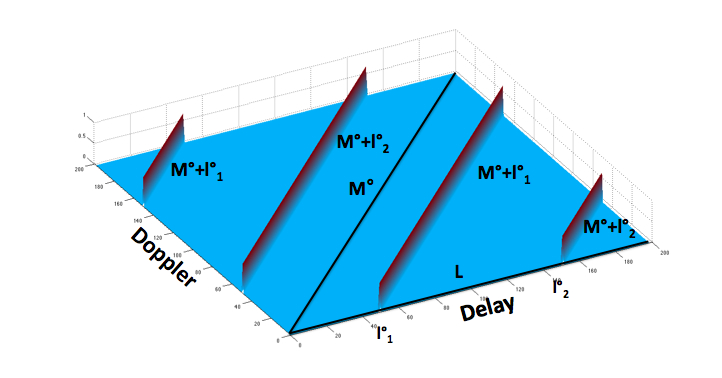}\\
\caption{$\left\vert \mathcal{A(}%
C_{M^{\circ }},H(C_{M^{\circ }}\mathcal{)})\right\vert $\textbf{\ }with%
\textbf{\ }$M^{\circ }=\{(\protect\tau ,\protect\tau )\},$\textbf{\ }and $%
supp(H)$ as in Figures \protect\ref{ChirpL}, and \protect\ref{ChirpM}. }
\label{ChirpN}
\end{figure}

We use a third chirp $C_{M^{\circ }},$ associated with a third line $%
M^{\circ }\subset V.$ Assume that $H$ is also $M^{\circ }$-generic. Then we
have---see Figure \ref{ChirpN} for illustration---that $\mathcal{A}%
(C_{M^{\circ }},H(C_{M^{\circ }}))$ is supported on $r$ shifted lines, which
are parallel to $M^{\circ }$ and pass through the delay-Doppler points that
we want to detect. To find these shifted lines we compute---see Figure \ref%
{ChirpN}---$\mathcal{A}(C_{M^{\circ }},H(C_{M^{\circ }}))$ on $L,$ and
obtain $r$ peaks at points 
\begin{equation}
l_{1}^{\circ },...,l_{r}^{\circ }\in L,  \label{l-dot-s}
\end{equation}%
yielding the $r$ desired shifted lines $M^{\circ }+l_{k}^{\circ },$ $%
k=1,...,r.$

 Under certain additional
genericity assumption (see Section \ref{Perf}) the solution to Problem \ref%
{M} is exactly the collection of \ the points $v_{i,j}=l_{i}+m_{j}$, $1\leq
i,j\leq r$, with three shifted lines from $L+m_{j}$'s$,$ $M+l_{i}$'s$,$ $%
M^{\circ }+l_{k}^{\circ }$'s passing through them---see Figure \ref%
{Incidence} for illustration. We summarize the computational part of the
method in the \textbf{Incidence Algorithm} below.\medskip

\begin{algorithm}
\underline{\textbf{Incidence Algorithm}}\textbf{\bigskip }

\begin{description}
\item[\textbf{Input:}]$\,$Chirps $C_{L},C_{M},C_{M^{\circ }}$, associated with randomly chosen lines $L,M,$ and $M^{\circ},$ and corresponding
echoes $R_{L},R_{M},R_{M^{\circ }}$, threshold $T > 0$, and value of $SNR$.

\medskip 

\item[\textbf{Output:}] $\,\,\,$ Channel parameters.

\end{description}

\medskip

\begin{enumerate}

\item Compute $\mathcal{A(}C_{M},R_{M}\mathcal{)}$ on $L$, obtain peaks at $l_{1},...,l_{r_1}. \smallskip $

\item Compute $\mathcal{A(}C_{L},R_{L}\mathcal{)}$ on $M,$ obtain peaks\footnote{We say that at  
$v \in V$ the ambiguity function of $C$ and $R$ has \textit{peak} if 
$|\mathcal{A}(C_L,R_L)[v]| \ge T \frac{\sqrt{2\log \log(N)}}{\sqrt{N \cdot SNR}}$} at $m_{1},...,m_{r_2}.\smallskip $

\item Compute $\mathcal{A(}C_{M^{\circ }},R_{M^{\circ }}\mathcal{)}$ on $L,$
obtain peaks at $l_{1}^{\circ },...,l_{r_3}^{\circ }.\smallskip $

\item Return the points $v_{i,j} \in V$ that satisfy $v_{i,j}=l_{i}+m_{j}\in M^{\circ }+l_{k}^{\circ },$ $1\leq i\leq r_1$, $1\leq j\leq r_2$, $1\leq k\leq r_3$. 
\end{enumerate}
\end{algorithm}\footnotetext{%
We say that at $v\in V$ the ambiguity function of $f$ and $g$ has \textit{%
peak,} if $\left\vert \mathcal{A}(f,g)[v]\right\vert >T\sqrt{2\log \log N}/%
\sqrt{N\cdot SNR}.$}

\begin{remark}[\textbf{Single Transmission}]
The incidence method can be modified to work with a single transmission, for
sensing targets with slightly larger attenuation coefficients. Indeed,
consider the sequence $C_{L,M,M^{\circ }}=\left( C_{L}+C_{M}+C_{M^{\circ
}}\right) /\sqrt{3}$\ constructed using chirps associated with three different
lines. Using Theorem \ref{CT}, and the normalization $\sum_{k=1}^{r}\left%
\vert \alpha _{k}\right\vert ^{2}\leq 1,$ we have 
\begin{equation*}
\mathcal{A}(C_{K},H(C_{L,M,M^{\circ }}))=\mathcal{A}(C_{K},H(C_{K}))/\sqrt{3}%
+O(\sqrt{r/N}),
\end{equation*}%
where $K=L,M$ or $M^{\circ }.$ In particular, the incidence algorithm above
is applicable.
\end{remark}

\begin{figure}[ht]
\includegraphics[clip,height=5cm]{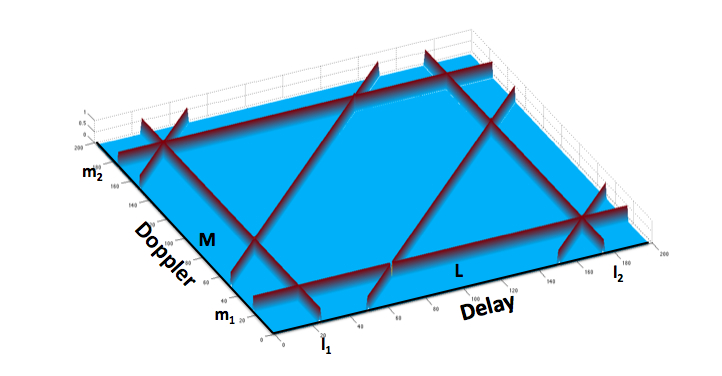}\\
\caption{Three chirps are used to obtain correct matching: $l_{1}$
with $m_{2},$ and $l_{2}$ with $m_{1}.$  }
\label{Incidence}
\end{figure}

\subsection{\textbf{Perfectness}\label{Perf}}

For the incidence method to work well we need additional relation, called 
\textit{perfectness}, between the channel operator $H$ and the lines $%
L,M,M^{\circ }.\medskip $

\begin{definition}
We define

\begin{enumerate}
\item Let $\mathcal{F}$ be a family of shifted lines in $V.$ A vector $v\in
V $ is called \textit{incidence point} of $\mathcal{F}$, if $v$ is lying on
at least two shifted lines from $\mathcal{F}$. The\textit{\ incidence number 
}of incidence point $v$ of $\mathcal{F}$, is the number of shifted lines
from $\mathcal{F}$ on which $v$ is lying. \medskip

\item A collection of vectors $v_{1},\ldots ,v_{r}\in V$ is called\textbf{\ }%
\textit{perfect} with respect to a collection of lines $L_{1},...,L_{d},$ if
these are the only incidence points of the family $\mathcal{F=}%
\{v_{i}+L_{j}\,|\,1\leq i\leq r,$ $1\leq j\leq d\}$ with incidence number $%
d.\medskip $
\end{enumerate}
\end{definition}

\begin{example}
In Figure \ref{Incidence} the vectors $l_{1}+m_{2}$, $l_{2}+m_{1},$ form a
perfect collection with respect to the lines $L,M,M^{\circ }.$ \medskip
\end{example}

Assume now that $r>0$ is given, and that $\#supp(H)=r$. Let us choose at
random three lines $L,M,M^{\circ }\subset V$. Assume that $H$ is generic
with respect to $L,M,$ and $M^{\circ }$. \medskip

\begin{proposition}[\textbf{Perfectness}]
\label{P}The probability $P$ that $supp(H)$ is perfect with respect to $%
L,M,M^{\circ }$ satisfies%
\begin{equation*}
P\geq 1-\frac{r(r^{2}-r)}{N}.
\end{equation*}
\end{proposition}

\subsection{\textbf{Remarks on Performance}}

\begin{enumerate}
\item \textbf{Detection. }In the noiseless scenario, it follows from Remark %
\ref{rG}, that in steps 1, 2, 3, of the incidence algorithm above, we have $%
r_{1}=r_{2}=r_{3}=r$, with probability greater or equal $1-O(r^{2}/N)$. In
addition, in this case it follows from Proposition \ref{P} that the
incidence method will return all the channel parameters with probability
greater or equal $1-O(r^{3}/N)$.\medskip 

\item \textbf{Noise. }Combining (\ref{noise}) and (\ref{AL}), we deduce that
in case of genericity and perfectness of $supp(H)$ with respect to the three
randomly chosen lines, the incidence algorithm will detect the delay-Doppler
shifts (\ref{DCP}) associated with attenuation coefficients of magnitude
larger then $(T+1)\sqrt{2\log \log N}/\sqrt{N\cdot SNR}$ with probability
going to one, as $N$ goes to infinity.\medskip

\item \textbf{Arithmetic Complexity. }Using Remark \ref{FC}, we can compute
all the $l_{i}$'s, $m_{j}$'s, and $l_{k}^{\circ }$'s, in $O(N\log N)$
operations. The verification of which of the $r^{2}$ points $l_{i}+m_{j},$ $%
1\leq i,j\leq r,$ lie on one of the shifted lines $M^{\circ }+l_{k}^{\circ }$
requires order of $r^{3}$ arithmetic operations. Overall, the arithmetic
complexity of the incidence method is $O(N\log N+r^{3}).\medskip $

\item \textbf{Real Time. }Applicability of incidence method for
inhomogeneous radar detection requires the transmission of three chirps. For
time-varying channel this might be not useful \cite{L}.\medskip
\end{enumerate}

\section{\textbf{The Cross Method\label{TCM}}}

The \textit{cross method} is the second resolution for the matching problem
that we discuss. We show how to use the values---including the phase and not
just the amplitude---of the ambiguity function, to suggest a solution to the
matching problem. This method does not require transmission of additional
chirp, and has lower arithmetic complexity.

\begin{figure}[ht]
\includegraphics[clip,height=5cm]{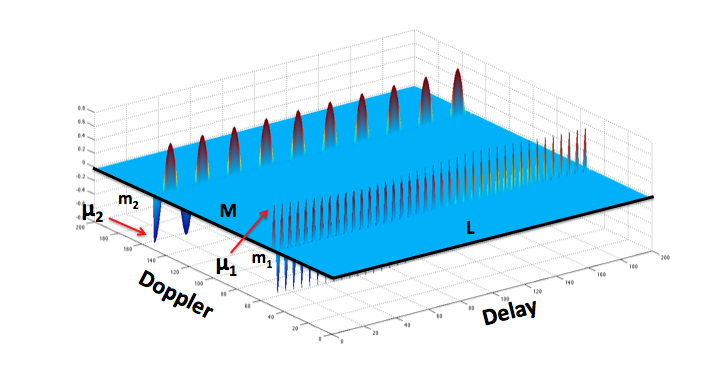}\\
\caption{The values $\protect\mu _{j}=\mathcal{A}(C_{L},H(C_{L}))[m_{j}],$ $%
j=1,2,$ are used for resolving the matching problem.  }
\label{Values-ChirpL}
\end{figure}


\begin{figure}[ht]
\includegraphics[clip,height=5cm]{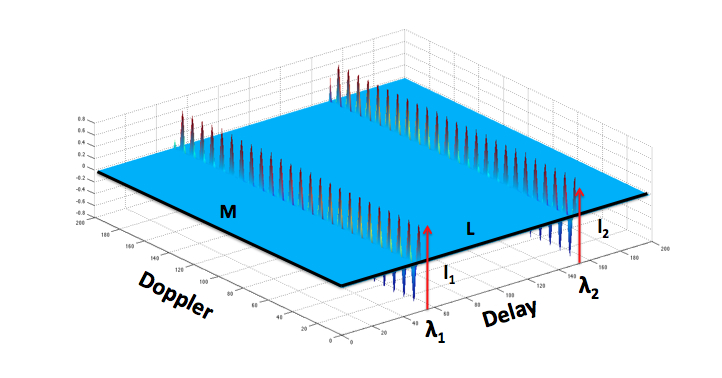}\\
\caption{The values $\protect\lambda _{i}=\mathcal{A}%
(C_{M},H(C_{M}))[l_{i}],$ $i=1,2,$ are used for resolving the matching
problem.  }
\label{Values-ChirpM}
\end{figure}


\subsection{\textbf{The Cross Method}}

Consider chirps $C_{L},C_{M},$ associated with the lines $L,M$ and
characters $\psi _{L}:L\rightarrow 
\mathbb{C}
^{\ast },$ $\psi _{M}:M\rightarrow 
\mathbb{C}
^{\ast }$, respectively. This means (see Section \ref{E}) that we have the
eigenfunction identities $\pi (l)C_{L}=\psi _{L}(l)C_{L}$, and $\pi
(m)C_{M}=\psi _{M}(m)C_{M},$ for every $l\in L,$ $m\in M.$ Let us assume
that the channel operator $H$ is generic with respect to $L$ and $M$. We
have---see Figures \ref{Values-ChirpL},\ref{Values-ChirpM} for
illustration---the peak values 
\begin{equation*}
\left\{ 
\begin{array}{c}
\mathcal{A}(C_{L},H(C_{L}))[m_{j}],\text{ \ }j=1,...,r; \\ 
\mathcal{A}(C_{M},H(C_{M}))[l_{i}],\text{ \ }i=1,...,r,%
\end{array}%
\right.
\end{equation*}%
where $m_{1},...,m_{r}\in M,$ $l_{1},...,l_{r}\in L,$ are the points given
by (\ref{ms}), and (\ref{ls}), respectively. To resolve the matching
problem, we define \textit{hypothesis} function $h:L\times M\rightarrow 
\mathbb{C}
$ by\medskip 
\begin{eqnarray}
h(l,m) &=&\mathcal{A}(C_{L},H(C_{L}))[m]\cdot \psi _{L}[l]  \label{h} \\
&&-\mathcal{A}(C_{M},H(C_{M}))[l]\cdot e(\Omega \lbrack l,m])\cdot \psi
_{M}[m],  \notag
\end{eqnarray}%
where\footnote{%
In linear algebra $\Omega $ is called \textit{symplectic form.}} $\Omega
:V\times V\rightarrow 
\mathbb{Z}
_{N}$ is given by $\Omega \lbrack (\tau ,\omega ),(\tau ^{\prime },\omega
^{\prime })]=\tau \omega ^{\prime }-\omega \tau ^{\prime }.\medskip $

\begin{theorem}[\textbf{Matching}]
\label{TM}Suppose $l_{i}+m_{j}\in supp(H),$ then $h(l_{i},m_{j})=0.\medskip $
\end{theorem}

\begin{remark}
The conclusion in Theorem \ref{TM} is not necessarily true if $H$ is not
generic with respect to $L$ or $M.\medskip $
\end{remark}

\begin{remark}[\textbf{Algebraic Genericity}]
Under natural (genericity) assumptions on the channel operator, and for
random choice of chirps, the "converse" of Theorem \ref{TM} is true with
high probability, i.e., if $h(l_{i},m_{j})$ $\approx 0,$ $1\leq i,j\leq r$,
then with high probability $l_{i}+m_{j}\in supp(H).$ \medskip A more precise
formulation and development of this statistical aspect will be published
elsewhere. 
\end{remark}

We summarize the computational part of the method in the \textbf{Cross
Algorithm }below\footnote{%
We update the hypothesis function to the noisy case, and set
\par
$h(l,m)=\mathcal{A}(C_{L},R_{L})[m]\cdot \psi _{L}[l]-\mathcal{A}%
(C_{M},R_{M})[l]\cdot e(\Omega \lbrack l,m])\cdot \psi _{M}[m].$}.

\begin{algorithm}
\underline{\textbf{Cross Algorithm}}\medskip 

\begin{description}

\item[\textbf{Input:}] $\,$Chirps $C_{L},C_{M}$, associated with randomly chosen lines 
$L, M$, and randomly chosen characters $\psi _{L},\psi _{M}$; corresponding echoes 
$R_{L},R_{M};$ thresholds $T_1, T_2 > 0$, and the value of $SNR$.

\medskip 

\item[\textbf{Output:}] $\,\,\,$ Channel parameters.

\end{description}

\medskip

\begin{enumerate}

\item Compute $\mathcal{A}(C_{M},R_{M})$ on $L,$ and take the $r_1$
peaks\footnote{We say that at $v \in V$ the ambiguity function of $f$ and $g$ has peak, if 
$\mathcal{A}(f,g)[v] \geq T \frac{\sqrt{\log \log (N)}}{\sqrt{N \cdot SNR}}$} located at points 
$l_{i},$ $1\leq i\leq r_1 \smallskip$.

\item \smallskip Compute $\mathcal{A}(C_{L},R_{L})$ on $M,$ and take the $r_2$
peaks located at the points $m_j, 1 \leq j \leq r_2 \smallskip$.

\item Return the points $v_{i,j}=l_{i}+m_{j}$, which solve 
$|h(l_{i},m_{j})| \leq T_2 \sqrt{2\log \log (N)}/\sqrt{N \cdot SNR}$, where $1\leq i\leq r_1, 1 \leq j \leq r_2$.
\end{enumerate}
\end{algorithm}\footnotetext{%
We say that at $v\in V$ the ambiguity function of $f$ and $g$ has \textit{%
peak}, if
\par
$\left\vert \mathcal{A}(f,g)[v]\right\vert >T_{1}\sqrt{2\log \log N}/\sqrt{%
N\cdot SNR}.$}

\begin{remark}[\textbf{Single Transmission}]
The cross method can be modified to work with a single transmission, for
sensing targets with slightly larger attenuation coefficients. Indeed,
consider the sequence $C_{L,M}=\left( C_{L}+C_{M}\right) /\sqrt{2}$
constructed using chirps associated with two different lines. Using Theorem %
\ref{CT}, and the normalization $\sum_{k=1}^{r}\left\vert \alpha
_{k}\right\vert ^{2}\leq 1,$ we have 
\begin{equation*}
\mathcal{A}(C_{K},H(C_{L,M}))=\mathcal{A}(C_{K},H(C_{K}))/\sqrt{2}+O(\sqrt{%
r/N}),
\end{equation*}%
where $K=L,M.$ In particular, the cross algorithm above is applicable.
\end{remark}

\subsection{\textbf{Remarks on Performance}}

We have

\begin{enumerate}
\item \textbf{Detection.  }In the noiseless scenario, it follows from Remark %
\ref{rG}, that in steps 1, 2, of the cross algorithm above, we have $%
r_{1}=r_{2}=r$, with probability greater or equal $1-O(r^{2}/N)$. In
addition, in this case it is not hard to see that the cross algorithm will
return all the channel parameters with probability greater or equal $%
1-O(r^{2}/N)$.\medskip 

\item \textbf{Arithmetic Complexity. }Using Remark \ref{FC}, we can compute
all the $l_{i}$'s, $m_{j}$'s, (\ref{ls}), (\ref{ms}), in $O(N\log N)$
operations. The computation for which of the $r^{2}$ pairs $l_{i},m_{j},$ $%
1\leq i,j\leq r,$ the hypothesis function $h(l_{i},m_{j})$ is sufficiently
small$,$ requires order of $r^{2}$ arithmetic operations. Overall, the
arithmetic complexity of the cross method is $O(N\log N+r^{2}).\medskip $

\item \textbf{Real Time. }Applicability of cross method for inhomogeneous
radar detection requires the transmission of two chirps.\medskip 
\end{enumerate}

\section{\textbf{Conclusions\label{Co}}}

\begin{figure}[ht]
\includegraphics[clip,height=2.4cm]{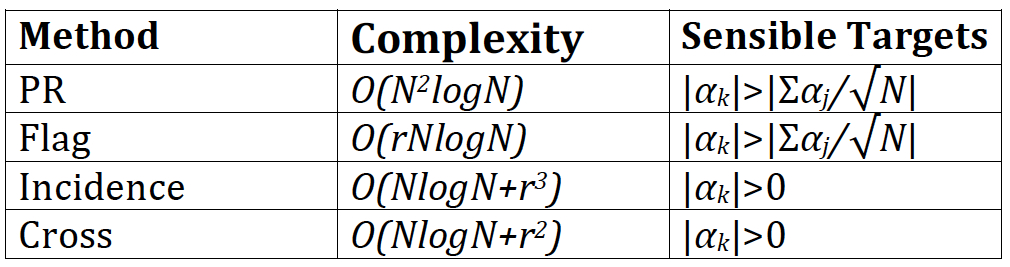}\\
\caption{Comparing methods, with respect to arithmetic complexity for $r$
targets, and sensibility of targets in terms of magnitude of attenuation
coefficients (noiseless scenario). }
\label{Table}
\end{figure}

In these notes we present the incidence and cross methods for efficient
radar detection. These methods, in particular, suggest solutions to two
important problems. The first is the inhomogeneous radar scene problem,
i.e., sensing small targets in the vicinity of large object. The second
problem is the arithmetic complexity problem. Low arithmetic complexity
enables higher velocity resolution of moving targets. We summarize these
important features in Figure \ref{Table}, and putting them in comparison
with the flag and pseudo-random (PR) methods.

\textbf{Acknowledgements. }We are grateful to our collaborators A. Sayeed,
and O. Schwartz, for many discussions related to the research reported in
these notes. We appreciate the contributions of I. Bilik, U. Mitra, K.
Scheim, and E. Weinstein, who shared with us some of their thoughts on radar
detection. We thank the students of the course "applied algebra", that took
place at UW\ - Madison in Fall 2013, and in particular we acknowledge the
Ph.D. students J. Lima, S. Qinyuan, and Z.M. Arslan. Finally, we thank the
Max Planck Institute for Mathematics at Bonn, and the Mathematics Department
at the Weizman Institute for Science, where part of this document was
drafted during July-August 2013.

\end{document}